\documentclass[aps,amsmath,amssymb,preprint]{revtex4}
\usepackage{graphicx}
\usepackage{dcolumn}
\usepackage{bm}
\usepackage{times}
\usepackage{amsmath}
\usepackage{epsfig}
\usepackage{feynmp}
\usepackage{epstopdf}
\begin{document}
\setlength{\unitlength}{1mm}
\title{Quantum criticality of U(1) gauge theories\\ with fermionic and bosonic matter in two spatial dimensions}
\author{Ribhu K. Kaul}
\affiliation{Department of Physics, Harvard University, Cambridge, MA 02138.}
\author{Subir Sachdev}
\affiliation{Department of Physics, Harvard University, Cambridge, MA 02138.}
\date{\today}
\begin{abstract}
We consider relativistic U(1) gauge theories in $2+1$ dimensions, with $N_b$ species of complex 
bosons and $N_f$ species of Dirac fermions at finite temperature. The quantum phase transition between the Higgs and Coulomb phases
is described by a conformal field theory (CFT).
At large $N_b$ and $N_f$, but for arbitrary values of the ratio $N_b/N_f$,  we present computations of various critical exponents and universal amplitudes for these CFTs. We make contact with the different spin-liquids, charge-liquids and deconfined critical points of quantum magnets that these field theories describe.  We compute physical observables that may be measured in experiments or numerical simulations of insulating and doped quantum magnets.  
\end{abstract}
\maketitle

\section{Introduction}

A number of experimental observations in the last two decades have begged for an understanding of interacting quantum systems that goes beyond the simple weakly interacting quasi-particle paradigm of solid-state physics. The large number of such systems with anomalous properties include as examples, cuprate superconductors, quantum critical heavy-fermion metals~\cite{stewart}, Mott-insulating organic materials~\cite{organics} and insulating frustrated magnets~\cite{kagome}. A new paradigm that has been invoked to describe these systems is quantum number fractionalization, see for e.g. Ref.~\onlinecite{lee}. Re-writing the original constituents in terms of fractional particles interacting with a gauge field, unconventional phases (or transitions) are accessed by the deconfined phases of these gauge theories. These deconfined phases cannot find a consistent description in terms of quasi-particles that are compositionally related to the  particles of the microscopic description, and hence provide a new paradigm that dramatically departs from the conventional confines of solid-state physics. 

An important issue that arises immediately is the detection of such exotic quantum phases or phase transition in experiments or numerical simulations. In most cases the appearance of these exotic field theories is also signaled by dramatic qualitative physical effects, e.g. a direct continuous transition between quantum states that break distinct symmetries~\cite{senthil1} (such a transition would not be permitted in the conventional Landau theory). While these novel physical effects themselves signal the appearance of the deconfined field theory, it is desirable to have a direct understanding and numerical estimate for universal quantities associated with the continuum field theories. An estimate for the universal numbers and scaling functions allows for a quantitative comparison between experiment/simulations and theory, and could be helpful for instance to distinguish between a non-universal weak first-order transition and a universal continuous one, in finite-size numerical computations~\cite{sandvik,mk,jiang}. 

The starting point of our analysis will be a relativistic continuum field theory of $N_f$ flavors of charged fermions and $N_b$ flavors of charged bosons mutually interacting through a gauge field in two spatial dimensions. This continuum description makes no reference to the variety of microscopic lattice models that may realize the quantum field theory. Examples of the derivations of relativistic field theoretic descriptions starting from lattice models of spins or electrons in different contexts may be found in Refs.~\onlinecite{rs,hermele,acl}. We postpone
a discussion of the application of our results to these physical realizations to Section~\ref{sec:conc}. 

We shall be interested in the general case of interacting field theories that involve bosons ($z_\alpha$), Dirac fermions ($\Psi_\alpha$) and gauge fields ($A_\mu$). Working in imaginary time the partition function of these theories may be written as functional integrals over the fields, 
\begin{equation}
Z = \int \mathcal{D} \Psi_\alpha \mathcal{D}z_\alpha\mathcal{D} A_\mu\exp \left( - S_{bf} \right),
\end{equation}
where the action $S_{bf} = S_b+S_f$ is specified below as the sum of bosonic and fermionic contributations.

In the field theory the bosons are described by the complex fields, $z_\alpha$, where $\alpha$ takes one of $N_b$ values. The $N_b$ fields couple through their mutual interaction with a `photon' gauge field, $A_\mu$, in the usual minimal coupling that preserves gauge invariance. 
\begin{equation}
\label{eq:bosonaction}
S_b =\frac{1}{g} \int d^2 r d\tau |(\partial_\mu - iA_\mu) z_\alpha|^2
\end{equation}  
The constraint $\sum_\alpha|z_\alpha|^2=1$ must be satisfied at all points in time and space. The action $S_b$ by itself is the well-known $\mathbb{CP}^{N_b-1}$ model. We remind the reader that in this theory, two-point correlation functions of $z_\alpha$ itself are 'gauge-dependent' quantities, but correlation functions of the composite operator $\phi^a=z^*_\alpha T^a_{\alpha\beta}z_\beta$ are not. Indeed, the `deconfined' theory of the N\'eel-VBS transition is described by $S_b$ at $N_b=2$, and $\phi^a$ plays the role of the N\'eel order parameter where $T^a_{\alpha\beta}$ are the generators of SU(2), a physical quantity that must be gauge independent. Some large-$N_b$ computations on this model are available in the literature~\cite{hlm,ikk}. See Ref.~\onlinecite{kleinert} for a review. We will recover these results as special cases of the more general field theory of interest here.

We now turn to the action that describes the Dirac fermions. The fermionic part is described by the usual Dirac fermion action,
\begin{equation}
\label{eq:fermionaction}
  S_f = \int d^2r d\tau \overline{\Psi}_\alpha[i\gamma^\mu(\partial_\mu-iA_\mu)] \Psi_\alpha
\end{equation}
where $\gamma^\mu$ satisfy the Clifford algebra in $2+1$ dimensions,
$\{\gamma^\mu,\gamma^\nu \}=2 \delta^{\mu\nu}$. We use the isotropic
version of $S_f$, since anisotropies have been shown to be irrelevant
at the critical fixed point studied here~\cite{vaftesfranz}. We shall consider the
action $S_b$ and $S_f$ separately and also $S_{bf}=S_b+S_f$ in the
large-$N_b,N_f$ limit. In general, a Maxwellian term for the $A_\mu$:
$S_A=\int d^2rd\tau (\epsilon_{\alpha\mu\nu}\partial_\mu A_\nu)^2$
should also be included. We note however that this term will not play
an important role in our considerations: As we will show below, the
contribution to the $A_\mu$ propagator from the interaction with
$z_\alpha$ and $\Psi_\alpha$ at large-$N$, decays with a slow power at
long distances and hence dominate the contribution from $S_A$ at large
length scales.  As is usual in quantum field theory, we introduce a
finite-$T$ by restricting the imaginary time integrals in the actions
above to extend from $0$ to $1/T$. The $z_\alpha$ and $A_\mu$ must
then satisfy periodic and the $\Psi_\alpha$ anti-periodic boundary
conditions.

We note here that the action $S_{ f}$, when considered separately is always quantum critical, i.e. there are no relevant perturbations consistent with its symmetry. This is in contrast to $S_{ b}$ which goes critical only at a single value of the 
coupling $g=g_c$, {\em i.e.\/} the boson mass is a relevant perturbation at the phase transition between the Higgs ($g<g_c$) and the deconfined Coulomb phase ($g>g_c$). The results in this paper include the results for $S_{\rm f}$ alone by simply setting $N_b=0$. Results for the $\mathbb{CP}^{N_b-1}$ model at criticality are obtained by setting $N_f=0$ and the general result for finite $N_b$ and $N_f$ applies for the Higgs to Coulomb phase transition in the background of critical Dirac fermion excitations.  We assume that $A_\mu$ is non-compact throughout this paper so that the 2+1-dimensional Coulomb phase is unequivocally stable (by definition) to monopole proliferation (see e.g. Ref.~\onlinecite{polyakov}).

The paper is organized in the following way: In Sec.~\ref{sec:S_eff} we first introduce the formal machinery of the large-$N$ limit of $S_{\rm bf}$ considered here (large-$N$ in this paper implies large-$N_b,N_f$ but an arbitrary finite value of the ratio). In Sec.~\ref{sec:expdim}, we then turn to an evaluation of the critical exponents and scaling dimensions of a number of quantities of physical interest at the $T=0$ critical point that separates the Higgs from the deconfined phase in the background of massless Dirac fermions. At the critical point of $S_{\rm bf}$, certain susceptibilities (corresponding to rotations in the boson or fermion flavor space) have simple linear or quadratic temperature dependencies with universal amplitudes because of conservation laws. In Sec.~\ref{sec:univamp}, we evaluate the numerical value of these amplitudes in the large-$N$ limit. Finally, in Sec.~\ref{sec:conc} we provide a general discussion of the physical situations in which the quantities computed in this publication may be measured in experiments or simulations of insulating and doped quantum anti-ferromagnets.

\section{Effective Action and Propogators at large-$N_b,N_f$}
\label{sec:S_eff}

In this section we provide the framework of the large-$N_b,N_f$ method used here and compute the effective action at large-$N$. At $N_b,N_f=\infty$, fluctuations of the gauge field are completely suppressed and we have a theory of free bosons and fermions. Gauge fluctuation at next-to-leading order are conveniently computed by perturbation theory, since each photon line brings a factor of $1/N$.   We hence require a computation of the photon propogator at order $1/N$, in the theory $S_{bf}$ introduced above at finite-$T$. In this paper we use two different gauges, depending on whether we are computing at $T=0$ or $T\neq 0$. At $T=0$, when space and time are completely interchangable, it is convenient to use a gauge that reflects this symmetry, we use $k_\mu A_\mu$ fixed to a constant. At $T\neq 0$, for practical computation reasons, we prefer to use the gauge $k_iA_i=0$ (with $i$ restricted to the two space indices). Throughout this paper, we will use greek symbols $\alpha,\beta,\mu,\nu$ etc. to take values $x,y$ and $\tau$ ($2+1$ dimensional space-time), and lower-case letters $i,j,k$ to take values $x,y$ ($2$-dimensional space). Summation is implied on repeated indices. 

\subsection{\label{sec:proptzero}Zero Temperature}

\begin{figure}
\includegraphics[width=2.7in]{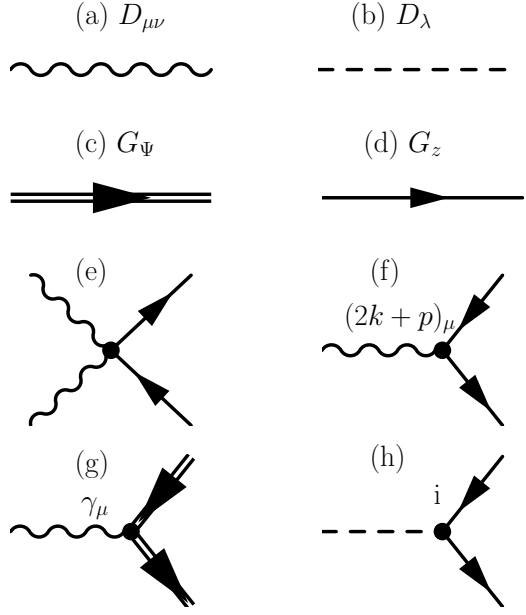}
\caption{ \label{fig:props}
Definition of diagrammatic symbols that arise from the action $S_b+S_f$. (a), (b), (c) and (d), illustrate the four propogators that are used to construct the various graphs used here. See Eq.~(\ref{eq:propzeroT}) for the $T=0$ values and Eqs.~(\ref{eq:propnzeroT1},\ref{eq:propnzeroT2}) for the $T\neq 0$. (e), (f), (g) and (h) show the four vertices allowed by our theory and their corresponding amplitudes. }
\end{figure}

We begin with the resolution of the constraint on $z_\alpha$, by introducing a real field $\lambda$, which acts as a Lagrange multiplier at each point of space and time. Including this into the bosonic part of the action, we can re-write it as,
\begin{equation}
S_b =\frac{1}{g} \int d^2 r d\tau \left [ |(\partial_\mu - iA_\mu) z_\alpha|^2 - i\lambda (|z_\alpha|^2-1)\right ]
\end{equation}  
Now after a standard sequence of steps (see e.g.~\cite{polyakov}) in the limit $N_b=\infty$ the gauge field drops out and $\lambda$ takes on a uniform saddle point value that extremizes the action $\lambda_0=ir$,
\begin{equation}
\int \frac{d^3p}{8\pi^3} \frac{1}{p^2+r} = \frac{1}{g}
\end{equation}
The propogator for the $z$ particles is thus $G_z^{-1}(k)=k^2+r$. At $N_b=\infty$, it becomes critical when $r=0$, i.e. at $g=g_c$ where $\frac{1}{g_c}=\int \frac{d^3p}{8\pi^3}$. 

The effective action is obtained to leading order in $N_b,N_f$ by integrating out the bosons and fermions perturbatively, i.e. by evaluating diagrams shown in Fig.~\ref{fig:diagseff}. At $T=0$, it is possible to obtain simple closed-form expressions for the effective action,
\begin{equation}
\label{eq:seffT0}
\mathcal{S}_{A-\lambda}=\int_p~~  
\frac{\Pi_\lambda}{2} \lambda^2 + \frac{\Pi_A}{2} (\delta_{\mu \nu}-\frac{p_\mu p_\nu}{p^2})A_\mu A_\nu 
\end{equation}
where
\begin{eqnarray}
\label{eq:piA_piL}
\Pi_\lambda(p,r)&=&N_b\frac{1}{4\pi p} \tan^{-1}\left (\frac{p}{2\sqrt{r}} \right)\nonumber\\
\Pi_A(p,r)&=& N_{f}\frac{ p}{16}\\ &+& N_b \left[\frac{p^2+4r}{8\pi p}\tan^{-1}\left (\frac{p}{2\sqrt{r}} \right )-\frac{\sqrt{r}}{4\pi}\right]\nonumber
\end{eqnarray}
Details of this computation are provided in Appendix~\ref{app:D1D2}. We note that $\Pi_\lambda(0,r)=(8\pi\sqrt{r})^{-1}$ and $\Pi_\lambda(0,0)=\int \frac{1}{p^4}$, which is formally IR divergent in $d=3$, but this will cancel out of all physical observables.

Using the gauge $k_\mu A_\mu=1-\zeta$ for this $T=0$ calculation, we find the following form for the propogators,
\begin{eqnarray}
\label{eq:propzeroT}
D_{\mu\nu} &=& \langle A_\mu A_\nu \rangle =\frac{1}{\Pi_A}(\delta_{\mu\nu}-\zeta \frac{q_\mu q_\nu}{q^2})\nonumber\\
D_\lambda &=&\langle \lambda \lambda \rangle= \frac{1}{\Pi_\lambda}\nonumber\\
G_z &=& \langle z^* z \rangle =\frac{1}{k^2 + r}\nonumber\\
G_\Psi &=& \frac{\not \! k}{k^2}
\end{eqnarray}
Where we have used the usual notation that $\not \! k = k^\mu\gamma^\mu$. The parameter $\zeta$ is an arbitrary gauge-fixing parameter and it should drop out of any expression for a physical observable.

\subsection{\label{sec:proptfin} Finite Temperatures}

At $T\neq 0$, there is no particular advantage in using the conventional relativistically invariant notation of Eq.~(\ref{eq:fermionaction}). So we use
the following equivalent form of the fermion action, designed for transparent appearance of frequency sums
\begin{eqnarray}
  S_f = \int d^2 x d \tau [ \Psi_\alpha^{\dagger} (\partial_\tau - i A_\tau  -i \sigma^x (\partial_x - i A_x )
      -i\sigma^y (\partial_y - i A_y) ) \Psi_\alpha ]
\end{eqnarray}
Again, the index $\alpha = 1 \ldots N_f$ labels the fermion flavors, 
$\Psi_\alpha$ is a
two-component Dirac spinor (for each $\alpha$), $\sigma^{x,y}$ are Pauli matrices acting on the Dirac space. 

\begin{figure}
\includegraphics[width=2.7in]{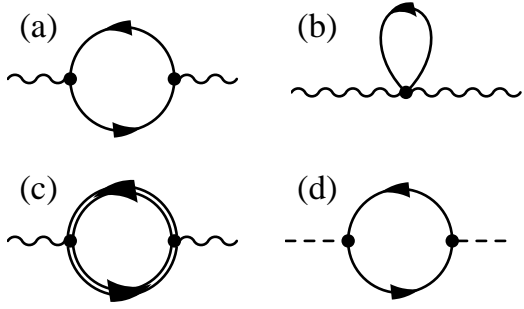}
\caption{\label{fig:diagseff} One-loop diagrams that determine the effective action at large-$N$. (a), (b) and (c) contribute to the effective action for the $A_\mu$ and (d) contributes to the action of the $\lambda$ field.}
\end{figure}

The general form of the effective action for the photon at large-$N_b,N_f$ can be simply written down by invoking a Ward identity,
\begin{widetext}
\begin{eqnarray}
\label{eq:SeffTfin}
\mathcal{S}_A = \frac{T}{2} \sum_{\epsilon_n} \int \frac{d^2 k}{4 \pi^2}
\left[ \left( k_i A_\tau - \epsilon_n A_i \right)^2 \frac{D_{1} ({\bf k},
\epsilon_n)}{{\bf k}^2} + A_i A_j \left( \delta_{ij} - \frac{k_i k_j}{{\bf k}^2}
\right) D_{2} ({\bf k}, \epsilon_n) \right].
\end{eqnarray}
\end{widetext}
where $D_1$ and $D_2$ are functions that can be evaluated at large-$N$ by perturbatively integrating out both the fermions and the bosons. This process is illustrated in the Feynman-graphs in Fig.~\ref{fig:diagseff}. A complete analytic evaluation at finite-$T$ of $D_{1,2}$ is not possible. However after some analytic manipulations detailed in Appendix~\ref{app:D1D2}, we can bring the expressions into forms which allow efficient numerical evaluation of these functions, these evaluations will play an important role later in our $T\neq 0$ computations. In terms of $D_{1,2}$, we can immediately write down the photon propogator in the Coulomb gauge $k_iA_i=0$.
 After imposing the gauge condition, the non-zero elements of the propagator are,
\begin{eqnarray}
\label{eq:propnzeroT1}
D_{00}({\bf q},\epsilon_n)&=&\frac{1}{D_1({\bf q},\epsilon_n)},\\
D_{ij}({\bf q},\epsilon_n)&=&\left( \delta_{ij} - \frac{q_i q_j}{{\bf q}^2} \right) \frac{1}{D_2({ \bf q},\epsilon_n) +
(\epsilon_n^2/{\bf q}^2) D_1({\bf q},\epsilon_n)}.\nonumber
\end{eqnarray}
The other propogators at finite-$T$ are,
\begin{eqnarray}
\label{eq:propnzeroT2}
G_z &=&\frac{1}{\nu_m^2+ {\bf k}^2 + r}\nonumber\\
G_\Psi &=& \frac{1}{-i\omega_n+\sigma \cdot {\bf k}}
\end{eqnarray}
where ${\bf k}$ is the the vector $(k_x,k_y)$ and $\nu_m = 2\pi T m$ and $\omega_n = 2\pi T (n+\frac{1}{2})$ are the usual bosonic and fermionic Matsubara frequencies.

\section{Critical Exponents and Scaling Dimensions at the QCP of $S_{bf}$}
\label{sec:expdim}

In this section we will work at $T=0$ and hence use the propogators derived in Sec.~\ref{sec:proptzero} and work in the gauge introduced there. As the coupling $g$ is tuned, a critical point at which the condensation of $z_\alpha$ takes place is crossed, for small $g$ the system enters a phase with condensed $z_\alpha$ and for large $g$ the system is described by a `liquid-like' state that breaks no symmetries (here the bosons $z_\alpha$ acquire a mass and the low-energy theory is a strongly coupled `algebraic' state of Dirac fermions coupled to a gauge field). In this section we first compute two independent critical exponents associated with the singularities at this transition ($\nu_N$ and $\eta_N$), including one-loop corrections about the  $N_f,N_b=\infty$ saddle point. We then turn to a calculation of the scaling dimension of the physical electron operator at the critical point.  

\subsection{Computation of $\nu_N$ in $S_{bf}$}
\label{sec:nuN}

\begin{figure}
\includegraphics[width=3in]{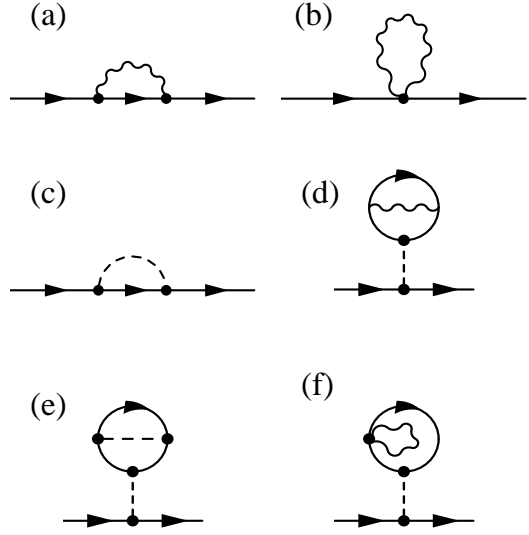}
\caption{ \label{fig:selfE}Self-energy diagrams that enter the two-point $z_\alpha$-boson correlation function at order $1/N$. These diagrams are evaluated in a specific gauge to extract the gauge-dependent indices $\eta_z$ and $\gamma_z$. A gauge independent index $\nu$ can then be extracted from these quantities by using the exponent relation, $\gamma_z = \nu_z (2-\eta_z)$.}
\end{figure}

We begin with the correlation length exponent, $\nu_N$,
\begin{equation}
\xi_N \propto (g-g_c)^{-\nu_N}, {~~~(\rm defines~~ } \nu_N) 
\end{equation}
where $\xi_N$ is the correlation length associated with the N\'eel order parameter. We note that in a given gauge one may calculate well-defined critical exponents using $z_\alpha$ as the order parameter. The critical indices so calculated are in general gauge-dependent. We note however that since the correlation length is gauge invariant so is the exponent $\nu_z$. Hence, $\nu_N$ may be calculated without worrying about composite operators, i.e. $\nu_N=\nu_z$. Other exponents must however be calculated using the composite N\'eel field, we will present calculations for $\eta_N$ in the following sub-section.

We calculate the exponent of interest, $\nu_z$, by calculating two gauge-dependent indices $\eta_z$ and $\gamma_z$ and then using the scaling relation $\gamma_z = \nu_z (2-\eta_z)$. Since $\nu_z$ should be gauge-independant, any dependence on the arbitrary gauge parameter, $\zeta$ should drop out of the result. This is an important check on our results. We begin by defining the anomalous dimension of the $z_\alpha$ field:
\begin{equation}
 {\rm dim} [z_\alpha]=\frac{D-2+\eta_z}{2}, {~~~(\rm in~ a ~given~ gauge,~defines~~ } \eta_z)
\end{equation}
We will make the gauge-dependence of $\eta_z$ explicit by calculating its value for arbitrary $\zeta$. The calculation of $\eta_z$ is straightforward and is obtained by picking up the co-efficient of the $p^2 \log p$ pieces in a perturbative expansion of $G_z$, and then re-exponentiating. Diagrams in Fig.~\ref{fig:selfE}(a,c) contribute giving,
\begin{equation}
\label{eq:eta_z}
\eta_z = \frac{4}{3\pi^2 N_b} - \frac{4}{(N_f + N_b)\pi^2}\left [\frac{10}{3} + 2 \zeta \right]
\end{equation}
For $\zeta=1$ and $N_f=0$, this reproduces the result of Ref.~\onlinecite{hlm}.

The critical index $\gamma_z$ determines how the mass of $z_\alpha$ (in a given gauge) goes to zero at the critical point, 
\begin{equation}
G_z^{-1}(k=0) = (g-g_c)^{\gamma_z}, {~~~(\rm defines~~ } \gamma_z)
\end{equation}
To leading order we may just use $r_c=\Sigma(0,0)$ and $\frac{1}{g}= \int \frac{1}{p^2+r}$, we find:
\begin{equation}
\frac{1}{g_c}= \int \frac{1}{p^2+r_c}= \int \frac{1}{p^2}- \int \frac{1}{p^4}\Sigma(0,0)
\end{equation}

Introduce a new parameter $r_g$, 
\begin{equation}
\frac{1}{g_c}-\frac{1}{g}= \int_p \frac{1}{p^2} - \frac{1}{p^2+r_g}= \frac{\sqrt{r_g}}{4\pi}
\end{equation}
which can be related to $r$ by,
\begin{eqnarray}
\frac{\sqrt{r_g}}{4\pi}&=&\frac{\sqrt{r}}{4\pi}-\int \frac{1}{p^4}\Sigma(0,0)\\
r &=& r_g  +\frac{\Pi_\lambda(0,0)}{\Pi_\lambda(0,r)}\Sigma(0,0)
\end{eqnarray}

Beyond the $N_b=\infty$ approximation the critical point is no more at $r_c=0$. In general,
\begin{equation}
\label{eq:zpropgen}
G^{-1}(p)= (p^2+r)-\Sigma(p,r)
\end{equation}
and so, $r_c=\Sigma(0,r_c)$. 

Now -writing the boson mass in terms of the new parameter $r_g$,
\begin{eqnarray}
\label{eq:ginv}
G^{-1}(0)= r_g - \left (\Sigma(0,r_g)- \frac{\Pi_\lambda(0,0)}{\Pi_\lambda(0,r)}\Sigma(0,0) \right),
\end{eqnarray}
we have eliminated the quantity $r$ completely. The goal now is to compute the term in brackets by evaluating the self-energy diagrams and pick up the terms with $r_g\log(r_g)$ divergences. Then these logs must be re-exponentiated yielding the exponent $\gamma_z$:
\begin{eqnarray}
G^{-1}(0)= r_g \left(1 + \alpha\log \left [\frac{\Lambda^2}{r_g}\right] \right)\approx |g-g_c|^{2(1-\alpha)}
\end{eqnarray}
where in the last equality we have used the fact that $r_g\propto |g-g_c|^2$, as derived above and that $\alpha$ vanishes in the $N\rightarrow \infty$ limit.

In order to demonstrate the usefulness of these formal manipulations, we calculate the index $\gamma_{\lambda}$ for the $O(N=2N_b)$ $\sigma$-model. This index is of course completely gauge invariant and follows by including only the $\lambda$ contributions to the self-energy, i.e. the diagrams in Figs.~\ref{fig:selfE}(c,e).
\begin{eqnarray}
\Sigma^{(c)}&=&\frac{2i^2}{2!}\int \frac{d^3q}{8\pi^3}\frac{1}{\Pi_\lambda(q,r)}\frac{1}{(p+q)^2 + r}\nonumber\\
\Sigma^{(e)}&=&\frac{4!i^4}{4!}\frac{1}{\Pi_\lambda(0,r)}\int \frac{d^3q}{8\pi^3}\frac{1}{(q^2+r)^2}\\
& &\times \int \frac{d^3p}{8\pi^3}\frac{1}{\Pi_\lambda(p,r)}\frac{1}{(p+q)^2 + r}
\end{eqnarray}
Now putting these expression back into Eq.~(\ref{eq:ginv}), we find that $\frac{\Pi_\lambda(0,0)}{\Pi_\lambda(0,r)}\Sigma(0,0)=0$. Note however that this term cannot be ignored in general. Indeed, when we include gauge fluctuations below, we find that it does not evaluate to zero. In the current case, because of the vanishing of this term, we find that the quantity in brackets in Eq.~(\ref{eq:ginv}) is simply $\Sigma^{(c)+(e)}(0,r)$,
\begin{eqnarray}
&=& - \int \frac{d^3p}{8\pi^3}\frac{1}{\Pi_\lambda(p,r)}\left [ \frac{1}{p^2+r} + \frac{\Pi^{\prime}_\lambda(p,r)}{2\Pi_\lambda(0,r)}\right]
\end{eqnarray}
where $\Pi_\lambda^\prime = \frac{\partial \Pi_\lambda}{\partial r}$. Now expanding the integrand for large $p$ we can extract the co-effecient $\alpha_\lambda=\frac{6}{\pi^2N_b}$. Thus $\gamma=2-\frac{12}{\pi^2 N_b}$, reproducing the standard result for the $O(N)$ model~\cite{skma}.

We now turn to the inclusion of the gauge field: There are four diagrams [Fig.~\ref{fig:selfE}(a,b,d,f)] that contribute to $\Sigma_A=\Sigma^{(a)} + \Sigma^{(b)} + \Sigma^{(d)} + \Sigma^{(f)}$. Including all factors of $i^2$, contractions and factorials, they are:
\begin{eqnarray}
\Sigma^{(a)}& =& \frac{2}{2!}\int \frac{d^3q}{8\pi^3}\frac{1}{\Pi_A(q,r)}\frac{N(p,q)}{(p+q)^2+r}\nonumber\\
\Sigma^{(b)} &=& (-3+\zeta)\int \frac{d^3q}{8\pi^3}\frac{1}{\Pi_A(q,r)}\nonumber\\
\Sigma^{(d)}&=& \frac{4 i^2}{2!~2!}\frac{1}{\Pi_\lambda(0,r)}\int \frac{d^3q}{8\pi^3}\frac{1}{\Pi_A(q,r)}\nonumber\\
& &\times \int \frac{d^3p}{8\pi^3} \frac{N(p,q)}{(p^2+r)^2((p+q)^2 + r)}\nonumber\\
\Sigma^{(f)}&=& \frac{(-3+\zeta)2 i^2}{2!~1!}\frac{1}{\Pi_\lambda(0,r)}\int \frac{d^3p}{8\pi^3}\frac{1}{(p^2+r)^2}\nonumber\\ 
& & \times \int \frac{d^3q}{8\pi^3}\frac{1}{\Pi_A(q,r)}
\end{eqnarray}
where $N(p,q)=4p^2-\frac{4\zeta}{q^2}(p\cdot q)^2 + (4p\cdot q + q^2)(1-\zeta)$, for a general gauge. Note that generally, $\Sigma^{(b)} +\Sigma^{(f)}=0$, so they never make an appearance. 

Our task is now to compute Eq.~(\ref{eq:ginv}) with $\Sigma=\Sigma^{(a)}+\Sigma^{(d)}$. This expression seems to be plagued with both IR and UV divergences, we will see below that both divergences exactly cancel as they must.
\begin{eqnarray}
\Sigma(0,r) &=& \int_q \frac{1}{\Pi_A(q,0)}\frac{q^2}{q^2+r}(1-\zeta)\nonumber\\ &-& \frac{1}{\Pi_\lambda(0,r)}\int_{p,q}\frac{N(p,q)}{(p^2+r)^2 [(p+q)^2+r]}\frac{1}{\Pi_A(q,r)}\nonumber\\
\Sigma(0,0) &=& \int_q \frac{1}{\Pi_A(q,0)}(1-\zeta)\nonumber\\ &-& \frac{1}{\Pi_\lambda(0,0)}\int_{p,q}\frac{N(p,q)}{p^4 (p+q)^2}\frac{1}{\Pi_A(q,0)}
\end{eqnarray}

Using these in Eq.~(\ref{eq:ginv}), we find for the term in brackets, $T_{(\cdots)}=\Sigma(0,r_g)- \frac{\Pi_\lambda(0,0)}{\Pi_\lambda(0,r_g)}\Sigma(0,0)$
\begin{eqnarray}
T_{(\cdots)} &=& (1-\zeta)\int_q \frac{1}{\Pi_A(q,r_g)}\frac{q^2}{q^2+r_g}\\
&-&\frac{1}{\Pi_\lambda(0,r_g)}\int_{p,q}\frac{N(p,q)}{(p^2+r_g)^2 [(p+q)^2+r_g]}\frac{1}{\Pi_A(q,r_g)}\nonumber\\
&+& \frac{1}{\Pi_\lambda(0,r_g)}\int \left[ \frac{N(p,q)}{(p+q)^2}-(1-\zeta)\right]\frac{1}{p^4}\frac{1}{\Pi_A(q,0)}\nonumber
\end{eqnarray}
Note that the last term was formed by combining two terms, both of which were IR divergent, this term however has no IR divergence. No other term has a IR divergence. We now proceed to show that all the UV divergences also exactly cancel. In order to do so, let us re-write,
\begin{eqnarray}
\label{eq:correction}
T_{(\cdots)} 
&=& (1-\zeta)\int_q \frac{1}{\Pi_A(q,r_g)} + (1-\zeta)\int_q \frac{1}{\Pi_A(q,r_g)}\frac{-r_g}{q^2+r_g}
\nonumber\\
&-&\frac{1}{\Pi_\lambda(0,r_g)}\int_{q}\frac{1}{\Pi_A(q,r_g)} I_A(q,r_g)\nonumber\\
&+& \frac{1}{\Pi_\lambda(0,r_g)}\int \frac{1}{4q}\frac{1}{\Pi_A(q,0)}
\end{eqnarray}
The following integral was used on the last line: $\int_p\left [\frac{N(p,q)}{(p+q)^2}-(1-\zeta)\right]\frac{1}{p^4}=\frac{1}{4q}$. 

Now, we can rewrite this integral as,
\begin{eqnarray}
\label{eq:correction}
T_{(\cdots)}&=&\int \frac{q^2 dq}{2\pi^2} \left [\frac{1-\zeta}{\Pi_A(q,r_g)}\frac{q^2}{q^2+r_g}\right.\\
&-&\left.\frac{1}{\Pi_\lambda(0,r_g)}\frac{I_A(q,r_g)}{\Pi_A(q,r_g)} + \frac{1}{\Pi_\lambda(0,r_g)} \frac{1}{4q}\frac{1}{\Pi_A(q,0)} \right ]\nonumber
\end{eqnarray}
All we have to do is to expand the term in $[...]$ for large q to extract the log divergence (Evaluation of $I_A$ is detailed in the Appendix~\ref{app:IA}. There should be a ${\rm O}(1/q^3)$ power but all lower powers must cancel. This is indeed found to be the case:
\begin{eqnarray}
T_{(\cdots)}&=& \int \frac{q^2 dq}{2\pi^2}\left [ 16 r_g \frac{7N_f-9N_b}{(N_b+N_f)^2}+\frac{\zeta}{N_f+N_b}\right ]\frac{1}{q^3}\nonumber\\
&=& -\alpha_A r_g\log \left[\frac{\Lambda^2}{r_g}\right]
\end{eqnarray}
where $\alpha_A=-\frac{4}{\pi^2}(\frac{7N_f-9N_b}{(N_b+N_f)^2}+\frac{\zeta}{N_f+N_b})$

Using $\alpha=\alpha_\lambda+\alpha_A$, and $\gamma=2(1-\alpha)$, we have the final expression for $\gamma$:
\begin{equation}
\label{eq:gamma}
\gamma= 2-\frac{12}{\pi^2 N_b} + \frac{8}{\pi^2}\left (\frac{7N_f-9N_b}{(N_b+N_f)^2}+\frac{\zeta}{N_f+N_b}\right )
\end{equation}
We note that when $\zeta=1, N_f=0$, this reduces to the result of Ref.~\onlinecite{hlm} (they only present results in the Landau gauge, $\zeta=1$). Using their quoted values of $\eta$ and $\nu$ and $\gamma=\nu (2-\eta)$, we find that their result gives (one has to also adjust $N$ by a factor of 2, because they have a theory with $N/2$ complex fields): $\gamma=2-76/(\pi^2 N_b)$, in agreement with us.

We can now calculate the co-efficient $\nu$ by using a scaling relation:
\begin{eqnarray}
\label{eq:nu_final}
\nu_z&=& \frac{\gamma_z}{2-\eta_z}\approx \frac{\gamma}{2}(1+\frac{\eta}{2})\\
    &\approx& \left [ 1-\frac{6}{\pi^2 N_b} + \frac{4}{\pi^2}\left (\frac{7N_f-9N_b}{(N_b+N_f)^2}+\frac{\zeta}{N_f+N_b}\right )\right ]\nonumber\\
 & &\times \left[ 1 +\frac{2}{3\pi^2 N_b} - \frac{2}{(N_f + N_b)\pi^2}[\frac{10}{3} + 2 \zeta]\right ]\nonumber\\
&\approx& 1 - \frac{16}{3\pi^2 N_b} + \frac{4}{\pi^2} \frac{7N_f-9N_b}{(N_b+N_f)^2} - \frac{20}{(N_f + N_b)3\pi^2}\nonumber
\end{eqnarray}
Reassuringly, the gauge parameter $\zeta$ drops out as expected. Also for $N_f=0$, $\nu = 1- \frac{48}{\pi^2 N_b}$, consistent with Ref.~\onlinecite{hlm}.

\subsection{Evaluation of $\eta_N$ in $S_{bf}$}

We now turn to an evaluation of the scaling dimension of the N\'eel field. We first define the anomalous dimension $\eta_N$ through the equation:
\begin{equation}
 {\rm dim} [\phi_N^a]=\frac{D-2+\eta_N}{2}
\end{equation}
where $\phi_N^a=z^*_\alpha T^a_{\alpha\beta}z_\beta$ is the N\'eel field and $T^a$ are the generators of SU$(N)$. Note that at $N_f,N_b=\infty$, the index $\eta_N=1$, i.e. it is already non-zero. This is unlike, e.g. the O(N) model, which at $N=\infty$ has no anomalous dimension. Since the N\'eel field is a composite of two $z_\alpha$ fields, its scaling dimension may be written as,
\begin{equation}
{\rm dim} [\phi_N^a]=2 {\rm dim}[z_\alpha] + \eta_{\rm vrtx}
\end{equation}
where $\eta_{\rm vrtx}$ gets contributions exclusively from vertex contributions illustrated in Fig.~\ref{fig:vertex}(a,b). This contribution is conveniently calculated by including a source $h_a \phi^a_N$ in the action and studying its renormalization, by momentum shell RG. There are two contributions coming from the photon and $\lambda$ propogators (integrals are from $\Lambda/s$ to $\Lambda$). They give:
\begin{eqnarray}
\delta h_a&=&-\frac{1}{2} 2i^2\int \frac{d^3q}{(2\pi)^3}\frac{1}{\Pi_\lambda}\frac{1}{q^4}  -\frac{1}{2}2\int \frac{d^3q}{(2\pi)^3}\frac{q^2(1-\zeta)}{\Pi_A}\frac{1}{q^4}\nonumber\\
& =& \frac{4}{\pi^2N_b} \log(s) - \frac{8}{\pi^2(N_f+N_b)}(1-\zeta) \log(s)
\end{eqnarray}
This gives $\eta_{\rm vrtx}= \frac{4}{\pi^2N_b} - \frac{8}{\pi^2(N_f+N_b)}(1-\zeta)$. We thus have the result,
\begin{eqnarray}
\label{eq:eta_final}
\eta_N &=& D-2 + 2\eta_z + 2\eta_{\rm vrtx}\\
&=& 1+ \frac{32}{3\pi^2 N_b} - \frac{128}{3\pi^2(N_f+N_b)}
\end{eqnarray}
Again $\zeta$ drops out as expected, since $\eta_N$ should be gauge invariant.

\subsection{Scaling Dimension of the Electron Operator at Criticality}

\begin{figure}
\includegraphics[width=2.in]{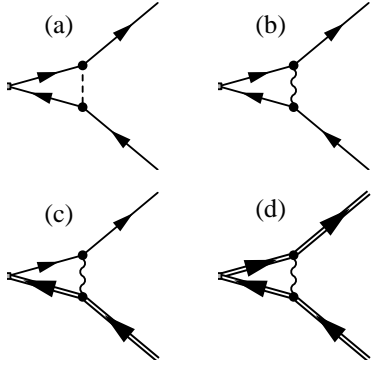}
\caption{ \label{fig:vertex} Vertex corrections to the scaling dimensions for the composite operators: (a,b) N\'eel vector, $\phi_N^a=z^*_\alpha T^a_{\alpha\beta}z_\beta$, (c) electron operator, $c=z^*\Psi$ and (d) fermion bilinear $\Psi^\dagger\Psi$}
\end{figure}

The scaling dimension of the physical electron operator is calculated at large-$N$ by identifying its scaling dimension with the bilinear $c = \overline{\Psi}z$. The scaling dimension of $c$ must clearly by gauge-invariant.

The scaling dimension of $\Psi$ is defined by:
\begin{equation}
{\rm dim}[\Psi] = \frac{D-1 + \eta_\Psi}{2}
\end{equation}

We can evaluate $\eta_\Psi$ by evaluating the lowest self-energy graph $\Sigma(k)$ due to the gauge field and then picking up the $\not \! k \log(k)$ part:
\begin{eqnarray}
\Sigma(k) &=& \frac{2}{2!} \int \frac{d^3q}{(2\pi)^3}\frac{1}{\Pi_A}\gamma_\mu \frac{\not \! q+\not \! k}{(k+q)^2}\gamma_\nu \left (\delta_{\mu\nu} - \zeta \frac{q_\mu q_\nu}{q^2}\right)\nonumber\\
&=& \left (\frac{8}{3\pi^2(N_f+N_b)}-\frac{8\zeta}{\pi^2(N_f+N_b)}\right) \not \! k \log(k)\nonumber
\end{eqnarray}
This gives the gauge-dependent result, $\eta_\Psi= \frac{8}{3\pi^2(N_b+N_f)} -\frac{8\zeta}{\pi^2(N_b+N_f)}$.

Now we can get the gauge invariant quantity, dim[c], by including the vertex contribution in Fig.~\ref{fig:vertex}(c).
\begin{eqnarray}
{\rm dim}[c] &=& {\rm dim}[z] + {\rm dim}[\Psi] + \eta_{\rm vrtx}\\
\end{eqnarray}
The parameter $\zeta$ drops out as expected. Evaluating the vertex contribution by the momentum shell method, we find,
\begin{eqnarray}
\eta_{\rm vrtx}\log(s) &=& \frac{-1}{1!1!} \int_{\Lambda/s}^\Lambda \frac{d^3q}{(2\pi)^3}\frac{1}{\Pi_A}\ q_\mu \frac{\not \! q}{q^4}\gamma_\nu \left (\delta_{\mu\nu} - \zeta \frac{q_\mu q_\nu}{q^2}\right)\nonumber \\ &=& -\frac{8}{\pi^2(N_f+N_b)}(1-\zeta)\log(s)
\end{eqnarray}
giving the final answer,
\begin{equation}
{\rm dim[c]}=\frac{3}{2} + \frac{2}{3\pi^2 N_b}-\frac{40}{3\pi^2 (N_b+N_f)}
\end{equation}

Finally we note that it is possible to construct yet another gauge invariant bilinear, $\Psi^\dagger\Psi$. Its scaling dimension, dim[${\Psi^\dagger} \Psi$], can again calculated for arbitrary $\zeta$, we find
\begin{equation}
{\rm dim}[{\Psi}^\dagger\Psi]= 2 -\frac{64}{3\pi^2 (N_b+N_f)}
\end{equation}
This quantity too gets a vertex contribution, Fig.~\ref{fig:vertex}(d), in addition to the $2{\rm dim}[\Psi]$ part. Again the full result is reassuringly gauge-invariant and for $N_b=0$ reproduces the results of Ref.~\onlinecite{rw} who calculated this number in the context of the `algebraic spin liquid'.

\section{Universal amplitudes of Susceptibilities and specific heat at finite-$T$ from Free Energy}
\label{sec:univamp}

In the previous section we studied the action $S_{\rm bf}$ at $T=0$, which corresponded to an infinite 3-dimensional system. We now turn to a study of some properties at $T\neq 0$, which corresponds to studying the field theory in a slab geometry in which two dimensions are infinite, but the third direction has a finite extent of $1/T$. This naturally complicates the computations, since the relativistic invariance of $T=0$ is destroyed and frequency integrals are replaced by discrete sums on Matsubara modes. In this section we evaluate the temperature dependence of the specific heat and certain susceptibilities that correspond to rotations among the fermionic and bosonic fields respectively. All quantities discussed in this section may be computed exactly from the low-temperature, low-magnetic field (to be defined below) expansion of the free-energy. We compute such an expansion in a the large $N_b$ and $N_f$ limit, allowing for an arbitrary value of the ratio, $N_b/N_f$. Since the computations are done at finite-$T$ we will work in the gauge introduced in Sec.~\ref{sec:proptfin} and use the form of the propogators derived there.

We shall consider two imaginary ``magnetic fields'', $H_b$ and $H_f$ that are applied in the ``direction'' parallel to the $SU(N)$ generator, diag(1 1 1 ... -1 -1 -1...). The chief advantage of using this generator is that it modifies the action in a very simple way. For the bosons, it shifts $\omega_n\rightarrow \omega_n - \theta H_b/2$, 
where $\theta=1$ for the first half of the components of $z_\alpha$ and $\theta=-1$ for the others. For the fermions it does the same, with $\omega_n\rightarrow \omega_n - \theta H_f$.

Once we have the free-energy as a function of $H_b,H_f$ and $T$, we can extract the specific heat and susceptibilities by differentiating,
\begin{eqnarray}
\chi_b & =& \frac{\partial^2\mathcal{F}}{\partial H_b^2}\\
\chi_f &=& \frac{\partial^2\mathcal{F}}{\partial H_f^2}\\
C_V &=& - \frac{\partial}{\partial T}\left [ T^2 \frac{\partial}{\partial T} \frac{\mathcal{F}}{T} \right ]=-T  \frac{\partial^2\mathcal{F}}{\partial T^2} 
\end{eqnarray}

From the analysis of Ref.~\onlinecite{csy}, it is know that $\chi_b,\chi_f \propto T$ and $C_V\propto T^2$. These simple integer scaling powers are due to energy conservation and bosonic and fermioinic rotational symmetry. 
The proportionality constants of $C_V,\chi_b,$ and $\chi_f$ are universal amplitudes ($\mathcal{A}_{C_V}, \mathcal{A}_{\chi_b}$ and $\mathcal{A}_{\chi_f}$),  divided by the square of a non-universal velocity, $c$. Since $c$ is non-universal, it must be determined separately for each model. 
In the field theoretic analysis of this Section, we will set $c=1$, though its presence must be kept in mind for comparisons to simulations/experiments.

At large $N_b$ and $N_f$ we can expand the free energy, $\mathcal{F}=-T\ln Z$
\begin{equation}
\mathcal{F} = N_b f^{0b} + N_f f^{0f} + f^{1\lambda} + f^{1A}\left (\frac{N_b}{N_f}\right).
\end{equation}
Note that in the $N=\infty$ limit, the gauge field disappears and the bosons and fermions de-couple so they contribute separately; in next to leading order this is not true. Hence, $f^{0b}$ and $f^{0f}$ are independant of the ratio $N_b/N_f$, whereas $f^{1A}$ is a function of the ratio ($f^{1\lambda}$ can only depend on $N_b$ and hence is also independent of the ratio $N_b/N_f$). In order to calculate the universal amplitudes associated with the specific heat and the susceptibilities we compute $f^{0b}$, $f^{0f}$, $f^{1\lambda}$ and $f^{1A}$ for arbitrary $T,H_b,H_f$ and $N_b/N_f$. 
We first outline the computation of $f^{0b},f^{0f}$ and $f^{1}$, then we present the numerical results for each of $\chi_b,\chi_f$ and $C_V$, as the ratio $N_b/N_f$ is varied.

When $N_b,N_f=\infty$, we only need to consider a simple gaussian theory of bosons and fermions and the computation of the free energy is straightforward. Including the shift in $\omega_n$ as discussed above, we obtain for $f^{0b}$ and $f^{0f}$,
\begin{widetext}
\begin{eqnarray}
\label{eq:free_en0}
f^{0b}& =& \frac{T}{2} \sum_{\omega_n} \int \frac{d^2 k}{4 \pi^2} \left[
\ln (k^2 + (\omega_n+H_b/2)^2 + m^2) +  \ln (k^2 + (\omega_n -  H_b/2)^2 + m^2) \right] - \frac{m^2}{g}\nonumber\\
f^{0f} &=& - T \int \frac{d^2 k}{4 \pi^2} \left[ \ln (1 + e^{-(k+iH_f)/T}) + \ln (1 + e^{-(k-iH_f)/T}) \right]
\end{eqnarray}
\end{widetext}
 where the mass, $m$ of the bosons which at the critical coupling depends only on $T$ and the applied field $H_b$. It can be computed by minimizing the free energy at $N_b=\infty$. This leads to the following non-linear equation.
\begin{equation}
T \sum_{\omega_n} \int \frac{d^2 k}{4 \pi^2} \left[ \frac{1}{2} \frac{1}{k^2 + (\omega_n +  \theta H_b/2)^2 + m^2}  
\right] = \int \frac{d^3 p}{8 \pi^3} \frac{1}{p^2}
\label{const}
\end{equation}
where $\theta$ should be summed over $\pm1$. Applying the Poisson summation formula and evaluating the integrals and sums, we have the equation for $m$,
\begin{equation}
\int \frac{d^2 k}{4 \pi^2} \frac{1}{2\sqrt{k^2 + m^2}} \left[ \frac{1}{e^{(\sqrt{k^2 + m^2}+i\theta H_b/2)/T} - 1}
\right] = \frac{m}{4 \pi}
\end{equation}
which has the solution
\begin{equation}
m = T\ln \left [ \frac{2\cos(H_b/2T)+1+\sqrt{(2\cos(H_b/2T)+1)^2-4}}{2} \right ]
\label{mh2}
\end{equation}
Computation of the various quantities of interest at $N=\infty$ follows simply by differentiating the free energy. 

Turning now to the $1/N$ correction to the free energy, $f_1$ receives two contributions, one from the Gaussian  fluctuations of the $\lambda$ field and the other from the Gaussian fluctuations of $A_\mu$. The result is,
\begin{eqnarray}
\label{eq:free_en1}
f^{1\lambda} &=& \frac{T}{2}\sum_n \int \frac{d^2k}{4\pi^2}  \ln\left (\Pi_\lambda  \right ) \\
f^{1A} &=& \frac{T}{2}\sum_n \int \frac{d^2k}{4\pi^2}  \ln\left (D_1\left[ D_2+\frac{\epsilon_n^2}{k^2}D_1 \right] \right ) 
\end{eqnarray}
where $\Pi_\lambda,D_1$ and $D_2$ are evaluated at $(k,\epsilon_n)$, and it must be kept in mind that they depend on $H_b,H_f$ through the shift of all internal boson and fermion frequencies (as detailed above)  in the evaluation of the one loop graphs for the $\lambda$ and $A_\mu$ propogators. We note that the value of $g_c$, gets a correction from its $N=\infty$ value, $\frac{1}{g_c}=\int \frac{d^3p}{8\pi^3}\frac{1}{p^2}$ (see Sec.~\ref{sec:proptzero}), which can be evaluated from Eq.~(\ref{eq:zpropgen}) to be $\delta \frac{1}{g_c}=\frac{4}{N_f+N_b}\int \frac{d^3q}{8\pi^3}\frac{1}{q^2}$, this shift which is cut-off dependent must be included in the free-energy to properly cancel all the UV diveregences at $\mathcal{O}(1/N)$ and produce the universal cutoff-independent results described below.
 
We now turn to a presentation of our results for $C_V,\chi_b$ and $\chi_f$, obtained by numerically differentiating computations of the free energy formulae outlined above.

\subsection{Specific Heat}

\begin{figure}[]
\includegraphics[width=3.2in]{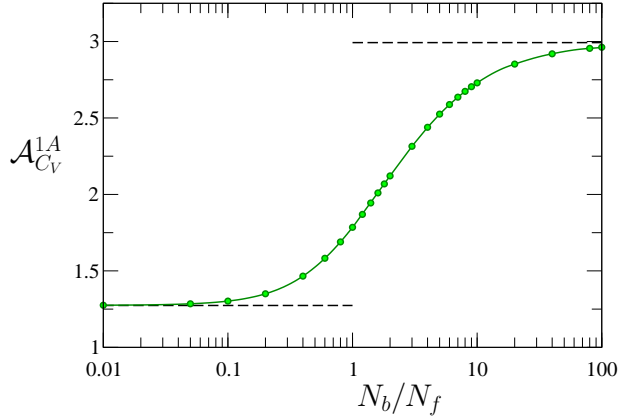}
\caption{ \label{fig:spheat} Gauge field fluctuation contribution to the specific heat amplitude, $\mathcal{A}^{1A}_{C_V}$ plotted as a function $N_b/N_f$. The values at $N_b/N_f=0,\infty$ are shown for reference as dashed lines.}
\end{figure}

The quantum critical specific heat at low-$T$ is of the form  $C_V = \mathcal{A}_{C_V} T^2$, where we can organize the large-$N$ expansion of the universal amplitude, $\mathcal{A}_{C_V}$ as,
\begin{equation}
\label{eq:spheat}
\mathcal{A}_{C_V}=N_b\mathcal{A}^{0b}_{C_V}+ N_f\mathcal{A}^{0f}_{C_V} + \mathcal{A}^{1\lambda}_{C_V} + \mathcal{A}^{1A}_{C_V}(N_b/N_f) 
\end{equation}
Differentiating Eq.~(\ref{eq:free_en0}), we obtain the estimates, $\mathcal{A}^{0b}_{C_V}=6 \frac{8\zeta(3)}{10\pi}\approx 1.83661$ and $\mathcal{A}^{0f}_{C_V}= 6 \frac{3\zeta(3)}{4\pi} \approx 1.72182$. Note that the proportionality of $\mathcal{A}^{0b}_{C_V}$ to $\zeta(3)$ with a rational co-efficient is non-trivial~\cite{polylog}.

Turning to the $1/N$ corrections, the computation of $\mathcal{A}^{1\lambda}_{C_V}$ and $\mathcal{A}^{1A}_{C_V}$ are far more complicated and involve some tedious numerical calculations. Basically numerical computations of the finite temperature propogators (as detailed in Appendix~\ref{app:D1D2}) have to be inserted into Eq.~(\ref{eq:free_en1}) which is then itself evaluated numerically. From our numerical analysis, we obtain, $\frac{T}{2}\sum_n \int \frac{d^2k}{4\pi^2}\ln \left (8\sqrt{k^2+\epsilon_n^2}\Pi(k,\epsilon_n) \right )=-0.03171 T^3$, allowing us to estimate, $\mathcal{A}^{1\lambda}_{C_V}=-0.38368$. We can get $\mathcal{A}^{1A}_{C_V}$ from a similar analysis, we plot it as a function of $N_b/N_f$ in Fig.~\ref{fig:spheat}.

\subsection{Susceptibility to applied $H_b$}

\begin{figure}[]
\includegraphics[width=3.2in]{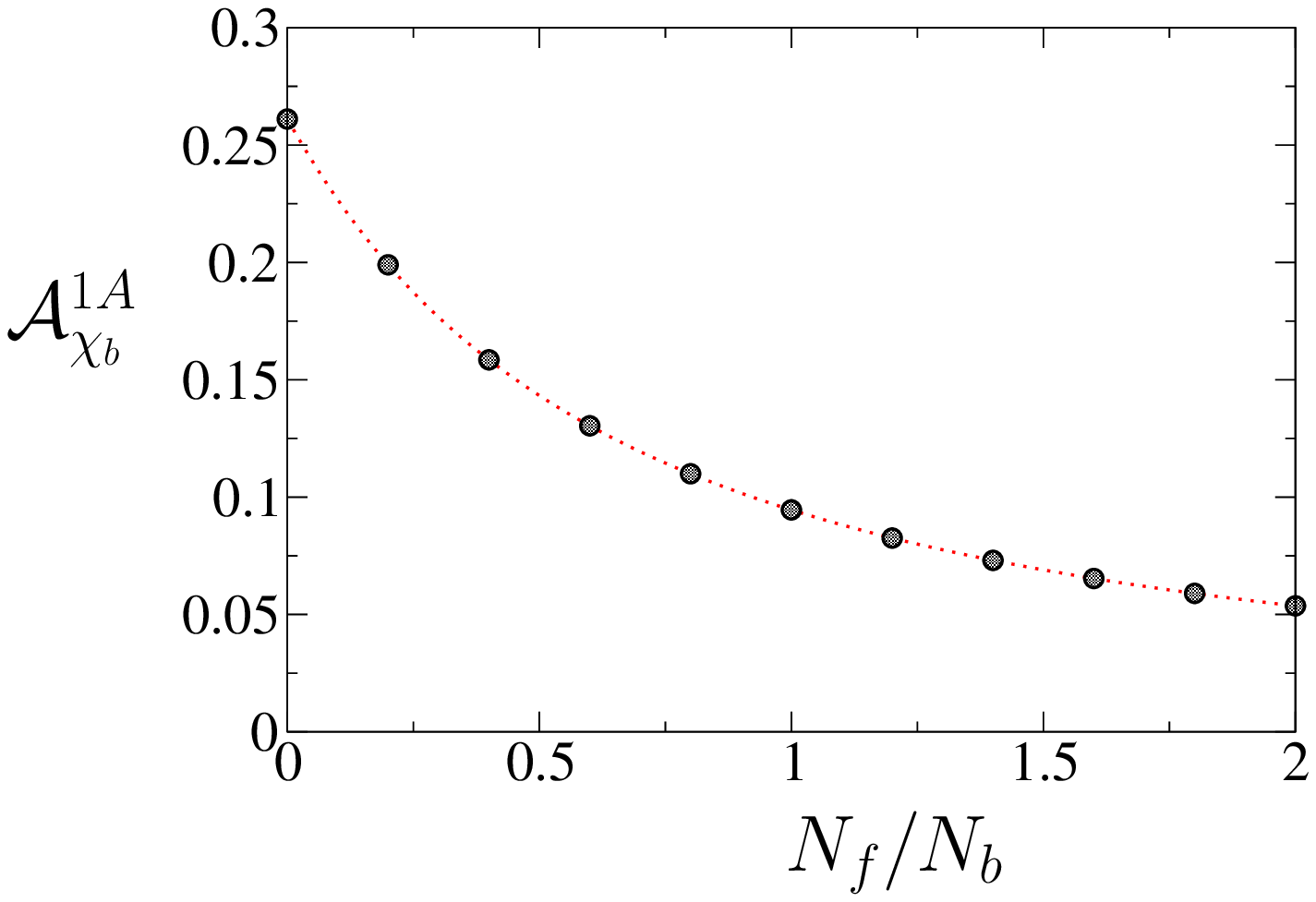}
\caption{ \label{fig:chib} Gauge field contribution to the amplitude of the susceptibility to applied $H_b$ as a function of $N_f/N_b$.}
\end{figure}

The susceptibility to $H_b$ is obtained by differentiating the free energy twice with respect to $H_b$ (keeping $H_f=0$ throughout). The large$-N$ expansion of the amplitude $\mathcal{A}_{\chi_b}$ defined from $\chi_b=\mathcal{A}_{\chi_b} T$ can be organized as,
\begin{equation}
\label{eq:chib}
\mathcal{A}_{\chi_b}=N_b\mathcal{A}^{0b}_{\chi_b} + \mathcal{A}^{1\lambda}_{\chi_b} + \mathcal{A}^{1A}_{\chi_b}(N_f/N_b) 
\end{equation}
We obtain the estimate $\mathcal{A}^{0b}_{\chi_b}=\frac{\sqrt{5}}{4\pi}\ln\left ( \frac{\sqrt{5}+1}{2}\right)\approx0.0856271$. From our numerical analysis described briefly above, we find  $\frac{T}{2}\sum_n \int \frac{d^2k}{4\pi^2}\ln \left (8\sqrt{k^2+\epsilon_n^2}\Pi(k,\epsilon_n) \right )=-0.03171 T^3-0.01325H_b^2T$. This allows us to estimate $\mathcal{A}^{1\lambda}_{\chi_b}\approx -0.02650$. Finally from the numerical evaluation of the gauge field, we plot the function $\mathcal{A}^{1A}_{\chi_f}(N_f/N_b)$ in Fig.~\ref{fig:chib}.
 
\subsection{Susceptibility to applied $H_f$}

\begin{figure}[]
\includegraphics[width=3.2in]{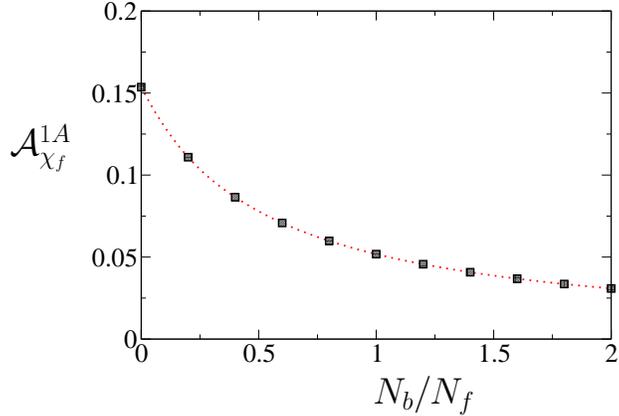}
\caption{ \label{fig:chif} Gauge field contribution to the amplitude of the susceptibility to applied $H_f$ as a function of $N_b/N_f$.}
\end{figure}

Finally, we turn to an evaluation of the susceptibility to $H_f$ which also has a linear-$T$ dependence. The universal amplitude appearing (just as in the case of $\chi_b$ above) can be expanded as,
\begin{equation}
\mathcal{A}_{\chi_f}=N_f\mathcal{A}^{0f}_{\chi_f} +  \mathcal{A}^{1A}_{\chi_f}(N_b/N_f) 
\end{equation}
In this case we obtain $\mathcal{A}^{0f}_{\chi_f}=2\frac{\ln(2)}{2\pi}\approx0.220636$. A computation of the function $\mathcal{A}^{1A}_{\chi_f}(N_b/N_f)$ along the lines described above is plotted in Fig.~\ref{fig:chif}.

\section{Conclusions}
\label{sec:conc}

All the computations carried out in this paper have been for the continuum field theory, $S_{\rm bf}$ and are hence of interest to any problem which realizes this field theoretic description. We now list three different physical problems of interest to which these results are applicable, and provide a brief summary of the connections between physical quantities and the universal numbers we have computed.

\textit{Deconfined N\'eel-VBS transition:} The deconfined N\'eel-VBS transition has been shown to be described by the field theory $S_{\rm bf}$ with $N_f=0$. The results of Sec.~\ref{sec:expdim} apply directly to the critical exponents of this transition (which has $N_f=0,N_b=2$). In particular, Eqs.~(\ref{eq:nu_final}) and (\ref{eq:eta_final}) characterize the two-point N\'eel correlation function. At the N\'eel-VBS quantum critical point, the uniform magnetic susceptibility, $\chi_u$ should depend linearly on temperature. This quantity is exactly the $\chi_b$, Eq.~(\ref{eq:chib}) computed in Sec.~\ref{sec:univamp} with $N_f=0$. The result for $C_V$, Eq.~(\ref{eq:spheat}) with $N_f=0$ also applies to the deconfined N\'eel-VBS transition. 

\textit{Algebraic Charge Liquids:} In a recent publication~\cite{acl}, a new class of states of matter, dubbed `algebraic charge liquids' (ACL) were shown to arise naturally in doped quantum anti-ferromagnets, which are close to the deconfined N\'eel-VBS quantum critical point. In the language of this study, $z_\alpha$ are bosons that carry spin and $\Psi$ are the superconducting quasi-particles. In this study the transition from a superconducting N\'eel state to the superconducting ACL state is described precisely by the field theory under study here with $N_b=2$ and $N_f=4$, whereas the criticality of the ACL state is described simply by $N_b=0, N_f=4$. The critical exponents of Sec.~\ref{sec:expdim} apply directly to the transition from the superconducting N\'eel state to the s-ACL state.  It has been shown~\cite{acl}, that the susceptibility $\chi_f$ discussed in Sec.~\ref{sec:univamp} is related to the finite-$T$ contribution to the superfluid density of the  s-ACL state (the result for $\chi_f$ at $N_b=0$ was published in Ref.~\onlinecite{acl}). Just like the N\'eel-VBS case the susceptibility $\chi_b$ is the uniform magnetic susceptibility, and in Sec.~\ref{sec:univamp}, we have the results for the universal amplitude both at the critical point and in the s-ACL phase. 

\textit{Algebraic Spin Liquids:} Finally the result for $\chi_f$ with $N_b=0$ applies to the fermionic spinon theories of algebraic spin liquids of square~\cite{lnw}, kagome~\cite{ran} and other lattices. In this case, $\chi_f$ is related to the uniform magnetic spin susceptibility because the fermions carry spin.

\section{Acknowledgements}

We acknowledge useful discussions with Y. Kim, T. Senthil and Z. Tesanovic. This research was supported 
by the NSF grants DMR-0537077 (SS and RKK), DMR-0132874 (RKK) and DMR-0541988 (RKK).

\appendix

\section{Effective Action at large-N: Evaluation of $D_1$, $D_2$, $\Pi_A$ and $\Pi_\lambda$}
\label{app:D1D2}

In this section we provide the details on the evaluation of the functions $D_1$ and $D_2$ that determine the effective action for the photons 
at large-$N$. We evaluate the general form of these functions at large $N_b,N_f$ and at finite-$T$, as defined in Eq.~(\ref{eq:SeffTfin}). From these results we can get $\Pi_A$  relevant to Sec.~\ref{sec:proptzero}, by setting $T=0$ and completing all integral analytically. Finally we provide a computation of the $\lambda$ propogator.

The evaluation of $D_1$ and $D_2$ corresponds to evaluating the diagrams in Fig.~\ref{fig:diagseff} at finite-$T$. Each functions gets a contribution from the bosonic as well as fermionic loops. We evaluated these separately, writing $D_1=D_{1b}+D_{1f}$ and $D_2=D_{2b}+D_{2f}$. We first present an evaluation of the fermionic loops and then turn to a computation of the bosonic contributions.

\subsection{$D_{1f}, D_{2f}$: Fermionic Loops}

The only diagram that needs to be evaluated for the fermions is the one appearing in Fig.~\ref{fig:diagseff}(c). Starting with the function $D_{1f} (k, \epsilon_n)$:
\begin{eqnarray}
&=& - N_f  \int_{{\bf q},\omega_n}
\mbox{Tr} \left[ 
G_\Psi ({\bf q},\omega_n)G_\Psi({\bf k+q},\omega_n + \epsilon_n)
\right]  \nonumber
\\
&=& 2 N_f \int_{{\bf q},\omega_n} 
\frac{\omega_n
(\omega_n+\epsilon_n) - {\bf q} \cdot ({\bf q} + {\bf
k})}{(\omega_n^2 + q^2)((\omega_n + \epsilon_n)^2 + ({\bf q} +
{\bf k})^2)} \\
&=& N_f \int_{{\bf q},\omega_n}
\left[\frac{-2}{q^2 + \omega_n^2} +  \frac{(2 \omega_n +\epsilon_n)^2
+k^2}{(\omega_n^2 + q^2)((\omega_n + \epsilon_n)^2 + ({\bf q} + {\bf
k})^2)} \right]\nonumber
\end{eqnarray}
We will use the shorthand $\int_{{\bf q},\omega_n} = T \sum_{\omega_n}
\int \frac{d^2 q}{4 \pi^2}$ to signify a summation on the internal
frequencies and momenta in this paper. Evaluating the first term by
$\zeta$-function regularization, and performing the momentum integration
in the second term using the usual Feynman trick, we find,
\begin{widetext}
\begin{eqnarray}
D_{1f} ({\bf k}, \epsilon_n) &=& \frac{N_fT \ln 2}{\pi}+
\frac{N_fT}{4 \pi}  \sum_{\omega_n}  \int_0^1 dx \frac{(2 \omega_n
+\epsilon_n)^2 +{\bf k}^2}{(1-x) \omega_n^2 + x (\omega_n
+ \epsilon_n)^2 + {\bf k}^2 x (1-x)} \nonumber \\
&=& \frac{N_fT \ln 2}{\pi}+ \frac{N_fT}{4 \pi}  \sum_{\omega_n}  2
\frac{(2 \omega_n +\epsilon_n)^2 +{\bf k}^2}{A} \ln \frac{A + {\bf k}^2 +
\omega_n^2 + (\omega_n + \epsilon_n)^2}{2|\omega_n (\omega_n +
\epsilon_n)|}\nonumber
\end{eqnarray}
where
\begin{equation}
A^2 = ({\bf k}^2 + (|\omega_n| - |\omega_n + \epsilon_n|)^2)({\bf k}^2 +
(|\omega_n| + |\omega_n + \epsilon_n|)^2).
\end{equation}
The frequency summation is formally divergent, but we can subtract
the divergent piece by zeta-function regularization to yield
\begin{eqnarray}
 D_{1f} ({\bf k}, \epsilon_n) &=& \frac{N_fT \ln 2}{\pi} \nonumber \\
&+&
\frac{N_fT}{4 \pi}  \sum_{\omega_n} \left\{ 2\frac{(2 \omega_n
+\epsilon_n)^2 +{\bf k}^2}{A} \ln \frac{A + {\bf k}^2 + \omega_n^2 + (\omega_n +
\epsilon_n)^2}{2|\omega_n (\omega_n + \epsilon_n)|} - 4 \right\}
\end{eqnarray}
\end{widetext}
This frequency summation is now evaluated numerically. For this, it
is useful to know the large $\omega_n$ behavior of the term in the
curly brackets. After symmetrizing the positive and negative
frequencies we obtain:
\begin{equation}
({\bf k}^2 + \epsilon_n^2) \left[ \frac{1}{3 \omega_n^2} +
\frac{9\epsilon_n^2 - {\bf k}^2}{30\omega_n^4} +\frac{50 \epsilon_n^4 -19
\epsilon_n^2 {\bf k}^2 + {\bf k}^4}{210\omega_n^6} + \ldots \right]
\end{equation}
The asymptotic behavior is then summed by using the identities
\begin{eqnarray}
\sum_{n=N+1}^{\infty} \frac{1}{(2n-1)^2} &=& \frac{1}{4N} -
\frac{1}{48N^3} + \frac{7}{960 N^5} + \ldots \nonumber \\
\sum_{n=N+1}^{\infty} \frac{1}{(2n-1)^4} &=&
\frac{1}{48N^3} - \frac{1}{96 N^5} + \ldots \nonumber \\
\sum_{n=N+1}^{\infty} \frac{1}{(2n-1)^6} &=& \frac{1}{320 N^5} +
\ldots
\end{eqnarray}

\begin{widetext}
We now turn to a similar evaluation as above for $D_{2f}$,
\begin{eqnarray}
D_{2f} (k, \epsilon_n) &=& -\frac{{\bf k}^2}{k_x k_y}
 N T \sum_{\omega_n} \int \frac{d^2 q}{4 \pi^2} \mbox{Tr} \left[
\sigma_x 
G_\Psi ({\bf q},\omega_n)
\sigma_y 
G_\Psi({\bf k+q},\omega_n + \epsilon_n)
\right] \nonumber
\\
&=&-\frac{{\bf k}^2}{k_x k_y} 2 N T \sum_{\omega_n} \int \frac{d^2 q}{4
\pi^2} \frac{2 q_x q_y + q_x k_y + q_y k_x}{(\omega_n^2 +
{\bf q}^2)((\omega_n + \epsilon_n)^2 + ({\bf q} + {\bf k})^2)} \nonumber
\\
&=& \frac{N_f T}{\pi} \int_0^1 dx \sum_{\omega_n}
\frac{x(1-x){\bf k}^2}{((1-x) \omega_n^2 + x (\omega_n + \epsilon_n)^2 +
  k^2 x(1-x))} \nonumber \\
&=& \frac{N_f T}{\pi} \sum_{\omega_n} \left\{ 1 +
\frac{(\omega_n+\epsilon_n)^2-\omega_n^2}{2{\bf k}^2} \ln
\frac{(\omega_n+\epsilon_n)^2}{\omega_n^2} \right. \nonumber \\
&-&
\frac{((\omega_n+\epsilon_n)^2 - \omega_n^2)^2 +
{\bf k}^2((\omega_n+\epsilon_n)^2 + \omega_n^2)}{A {\bf k}^2}  \nonumber
\\ &~&~~~~~~~~~~~~ \left. \times \ln  \frac{(A + {\bf k}^2 + \omega_n^2 + (\omega_n +
\epsilon_n)^2)}{2|\omega_n (\omega_n + \epsilon_n)|}  \right\}
\end{eqnarray}
\end{widetext}
The large $\omega_n$ behavior of the term in the curly brackets,
after symmetrizing the positive and negative frequencies, is:
\begin{equation}
{\bf k}^2 \left[ \frac{1 }{6 \omega_n^2} + \frac{7 \epsilon_n^2 -
2{\bf k}^2}{60\omega_n^4} +\frac{13 \epsilon_n^4 -34 \epsilon_n^2 {\bf k}^2 + 3
{\bf k}^4}{420\omega_n^6} + \ldots \right]
\end{equation}

At $T=0$, the exact values of $D_{1,2 f}$ can be computed analytically,
\begin{equation}
\label{eq:D12fT0}
D_{1f} ({\bf k}, \omega) = D_{2f} ({\bf k}, \omega) = \frac{N_f {\bf k}^2}{ 16\sqrt{{\bf k}^2 + \omega^2}}.
\end{equation}

\subsection{$D_{1b},D_{2b}$: Bosonic Loops}

We now present the evaluation of $D_{1b}$ and $D_{2b}$ that arise from the diagrams in Fig.~\ref{fig:diagseff}(a,b).
The function $D_{1b}$ is
\begin{widetext}
\begin{eqnarray}
D_{1b} ({\bf k}, \epsilon_n) &=&  N_b T \sum_{\omega_n} \int \frac{d^2
q}{4 \pi^2} \left[ \frac{2}{{\bf q}^2 + \omega_n^2+ r} \right.\nonumber \\
&-& \left.\frac{(2
\omega_n + \epsilon_n)^2}{({\bf q}^2 + \omega_n^2+ r)(({\bf q+k})^2 +
(\omega_n+\epsilon_n)^2+ r)} \right]
\end{eqnarray}
\end{widetext}
The first term we evaluate by dimensional regularization
\begin{eqnarray}
&=&T \sum_{\omega_n} \int \frac{d^d q}{4 \pi^2}  \frac{2}{{\bf q}^2 +
\omega_n^2+ r}\\
 &=& \frac{\Gamma(1-d/2)}{(4 \pi)^{d/2}} T
\sum_n \frac{1}{(\omega_n^2 + r)^{1-d/2}} \nonumber \\
&=& \frac{\Gamma(1-d/2)}{(4 \pi)^{d/2}} T^{d-1}
\left[ \left(\frac{T}{\sqrt{r}}\right)^{2-d} \right.\\
& &+ \left. (d-2) \sum_{n=1}^{\infty} \ln
\left( 1 + \frac{ r}{\omega_n^2} \right) + \frac{2}{(2
\pi)^{2-d}} \zeta(2-d) \right] \nonumber \\
&=& - \frac{1}{2 \pi} \left[ \ln \left( \frac{\sqrt{r}}{T} \right) +
\sum_{n=1}^{\infty} \ln \left( 1 + \frac{ r}{\omega_n^2}
\right)\right] \nonumber \\
&=& 0
\end{eqnarray}
where we have used the large-$N$ value for the mass parameter $\sqrt{r} = 2 \ln ((\sqrt{5}+1)/2)$ in the last line (see Ch. 5 of Ref.~\onlinecite{subirbook} for more details).

\begin{widetext}
\begin{eqnarray}
D_{1b} ({\bf k}, \epsilon_n) &=&  \frac{N_b T}{4 \pi} \sum_{\omega_n}
\int_0^1 dx  \left[ 4 - \frac{(2 \omega_n + \epsilon_n)^2}{(1-x)
\omega_n^2+ x (\omega_n+\epsilon_n)^2+ r + {\bf k}^2 x (1-x)}
\right].
\end{eqnarray}
The frequency summand has the following expansion at large
$\omega_n$
\begin{eqnarray}
&& \frac{ 2{\bf k}^2 + 12 r - \epsilon_n^2}{3 \omega_n^2} +
 \frac{-4 {\bf k}^4 + {\bf k}^2 (-40 r+ 17 \epsilon_n^2) - 3 (40 r^2 - 50 r \epsilon_n^2 + 3 \epsilon_n^4)}{30 \omega_n^4} \nonumber \\
 &&~+ \frac{6 {\bf k}^6+\left(84 r-73 \epsilon_n^2\right) {\bf k}^4+\left(420 r^2-826 \epsilon_n^2
   r+81 \epsilon_n ^4\right) {\bf k}^2+10 \left(84 r^3-315 \epsilon_n ^2 r^2+147 \epsilon_n ^4 r-5 \epsilon_n ^6\right)}{210\omega_n^6}\nonumber
\end{eqnarray}
For $D_{2b}$ we have
\begin{eqnarray}
D_{2b} ({\bf k}, \epsilon_n) &=&  \frac{N_b T}{4 \pi} \sum_{\omega_n}
\int_0^1 dx  \frac{{\bf k}^2 (2x-1)^2}{(1-x) \omega_n^2+ x
(\omega_n+\epsilon_n)^2+ r + {\bf k}^2 x (1-x)}.
\end{eqnarray}
and the following expansion at large $\omega_n$
\begin{equation}
{\bf k}^2 \left[ \frac{1}{3 \omega_n^2} -
 \frac{{\bf k}^2 + 10 r - 11 \epsilon_n^2}{30 \omega_n^4} + \frac{ {\bf k}^4 + 70 r^2
 -266 r \epsilon_n^2 + 86 \epsilon_n^4 + 14 {\bf k}^2 r - 23 {\bf k}^2
\epsilon_n^2 }{210 \omega_n^6} \right]
\end{equation}
\end{widetext}

\subsection{$\Pi_A$ and $\Pi_\lambda$: Action at $T=0$}

The above expressions for $D_1,D_2$ at finite-$T$ were brought into a form from which efficient numerical evaluation was possible. At $T=0$, a full analytic evaluation of the integrals entering the effective action is possible. We will outline the steps of these calculations now. We begin with a computation of $\Pi_A$. Just as in the finite-$T$ case, $\Pi_A= \Pi^b_A+\Pi_A^f$ receives contribution from both bosonic and fermionic loops, which follow from the $T\rightarrow 0$ limit of the above expressions for the finite-$T$ action. Using the form in Eq.~(\ref{eq:seffT0}), we can deduce by looking for example at the $q_x q_y$ term,
\begin{equation}
-\Pi^b_A \frac{q_x q_y}{q^2}= -N_b\int \frac{d^3q}{8\pi^3} \frac{(2k+q)_x (2k+q)_y}{(k^2+r)[(k+q)^2+r]}
\end{equation}
Now the evaluation of this integral follows the standard steps (outlined for example in Appendix~\ref{app:IA}) and we find,
\begin{equation}
\Pi^b_A(p,r)= N_b \left[\frac{p^2+4r}{8\pi p}\tan^{-1}\left (\frac{p}{2\sqrt{r}} \right )-\frac{\sqrt{r}}{4\pi}\right]
\end{equation}
The fermionic contribution can be evaluated in a similar way or by noting the simple relationship, $\Pi_A= \frac{{\bf p}^2+\omega^2}{{\bf p}^2}D_2({\bf p},\omega)$, giving from Eq.~\ref{eq:D12fT0} that,
\begin{equation}
\Pi^f_A(p) = \frac{N_f p }{16}
\end{equation}
Putting these results together we get the expression for $\Pi_A$ quoted in Eq.~(\ref{eq:piA_piL}). It is curious to note that when $r=0$ the bosonic contribution becomes $\Pi^b_A(p,0)=\frac{N_b p }{16}$, exactly the same as for the fermions.

The evaluation of $\Pi_\lambda$ follows the same steps as for $\Pi^b_A$ except with different contributions at the vertices, Fig.~\ref{fig:diagseff}(d). 
\begin{equation}
\Pi_\lambda(q) = N_b\int \frac{d^3 k}{8\pi^3}\frac{1}{(k^2+r)[(k+q)^2+r]}
\end{equation}
which can be evaluated using the usual Feynman parameter and integration shift of Appendix~\ref{app:IA}, resulting in the form quoted in Eq.~(\ref{eq:piA_piL}).

\section{Evaluation of $I_A$}
\label{app:IA}
Here we explicitly show how to calculate $I_A$ that appears in computations of Subsection~\ref{sec:nuN}. We note that we have used the standard sequence of steps illustrated in this example below many times in the course of the computations presented in this paper.
\begin{equation}
I_A=\int \frac{d^3p}{8\pi^3}\frac{4p^2-\frac{4\zeta}{q^2}(p\cdot q)^2 + (4p\cdot q + q^2)(1-\zeta)}{(p^2+r_g)^2 [(p+q)^2+r_g]}
\end{equation}
The integral is evaluated by executing the following steps in sequence: introduce a Feynman parameter $x$, shift to $l=p+xq$, integrate over $l$ and then integrate over $x$.
\begin{widetext}
\begin{eqnarray}
I_A&=&\int_0^1 dx~~ 2(1-x) \int \frac{d^3p}{8\pi^3}\frac{4p^2-\frac{4\zeta}{q^2}(p\cdot q)^2 + (4p\cdot q + q^2)(1-\zeta)}{((1-x)(p^2+r_g) + x[(p+q)^2+r_g])^3}\nonumber\\
&=&\int_0^1 dx~~ 2(1-x) \int \frac{d^3l}{8\pi^3}\frac{4l^2(1-\frac{\zeta}{3})+ q^2(1-\zeta)(4x(x-1)+1)}{(l^2 + r +q^2x(1-x))^3}\nonumber\\
&=&\int_0^1 dx 2 (1-x) \left [ \frac{3}{8\pi} \frac{1-\zeta/3}{\sqrt{q^2(1-x) x +r}} + \frac{[4x(x-1) +1] (1-\zeta ) q^2}{32 \pi }  \frac{1}{\left ( \sqrt{ q^2 (1-x) x+r} \right)^3 } 
\right ]\nonumber\\
&\approx& (1-\zeta)\Pi_\lambda(0,r_g)+\frac{1}{4q} -\frac{\sqrt{r_g}}{\pi q^2}+ {\rm O}(\frac{1}{q^3})
\end{eqnarray}
\end{widetext}
The neglected terms produce UV convergent diagrams when the integrals over $q$ are evaluated and hence can be neglected.

\section{Diagramatic Evaluation of Fermion Susceptibility}

In this Appendix we calculate one of the universal amplitudes of Sec.~\ref{sec:univamp}, $\mathcal{A}_{\chi_f}$, by an entirely different diagrammatic method. The perfect agreement between the two methods provides a non-trivial check on our rather involved computations. 

\begin{figure}
\includegraphics[width=2.7in]{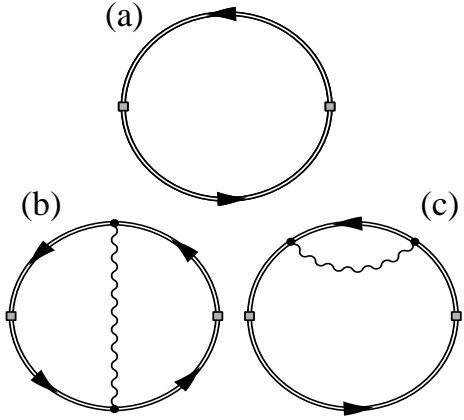}
\caption{ \label{fig:chiPsi} Diagrammatic contributions to the susceptibility, $\chi_f$ arising (a) at $N=\infty$, (b,c) at order $1/N$. All the diagrams give a $T$-linear dependence as expected for the susceptibility of a conserved charge for $z=1$ criticality in $d=2$.~\cite{csy}}
\end{figure}
Consider the susceptibility, $\chi_f$ of the SU$(N_f)$ charge, $Q^a=\Psi^\dagger_\alpha T^a_{\alpha\beta}\Psi_\beta$, where $T^a$ is a generator of the symmetry. Since the SU$(N_f)$ charge is a conserved density it acquires no anomalous dimension. Indeed a simple scaling analysis~\cite{csy} establishes that for $z=1$ criticality, in $d=2$ spatial dimensions, it must have a linear-$T$ dependence with a co-efficient that is a universal amplitude $\mathcal{A}_{\chi_f}$ divided by the square of a non-universal velocity. 
\begin{equation}
\chi_f = \frac{\mathcal{A}_{\chi_f}}{c^2}T
\end{equation}
We will set $c=1$ in the following. In general, we can think of the universal amplitude as a universal function $\mathcal{A}_{\chi_f}=\mathcal{A}_{\chi_f}(N_f,N_b)$ for arbitrary $N_f,N_b$. In the following we will calculate this functional dependence at large $N_f,N_b$ but for arbitrary values of the ratio $N_b/N_f$.

{\em $\mathcal{A}_{\chi_f}$ at $N=\infty$:} 
At $N
=\infty$, we have a free theory of Dirac fermions and $\chi_f$ can be computed by simply evaluating the diagram in Fig.~\ref{fig:chiPsi}(a).
\begin{eqnarray}
\chi_f &=& -\left(\mbox{Tr} (T^a)^2 \right) T \sum_{\omega_n} \int
\frac{d^2 k}{4 \pi^2} \mbox{Tr} \left(- i \omega_n + \vec{\sigma}
\cdot
{\bf k} \right)^{-2} \nonumber \\
&=& 2\left(\mbox{Tr} (T^a)^2 \right)  \sum_{\omega_n}\int \frac{d^2 k}{4 \pi^2}
\frac{\omega_n^2-k^2}{(k^2 + \omega_n^2)^2}
\end{eqnarray}
We will evaluate the momentum integration first, using dimensional
regularization, and then the frequency summation, using zeta
function regularization. It is easy to obtain the final answer by
evaluating the frequency summation before the momentum integration,
in which case no cutoff or regularization is needed. However, on including
gauge fluctuations, it appears more convenient to perform the momentum
integration first, this $N=\infty$ calculation then provides an opportunity to illustrate
the method by which we regulate the frequency sums later on. Performing the momentum integration in $d$
spatial dimensions, we obtain
\begin{equation}
\chi_f = \left(\mbox{Tr} (T^a)^2 \right) \frac{2(1-d)\Gamma(1-d/2)}{(4
\pi)^{d/2}} T \sum_{\omega_n} |\omega|^{d-2} \label{e1}
\end{equation}
We evaluate the frequency summation using
\begin{equation}
g(s) = T \sum_{\omega_n} \frac{1}{|\omega_n|^s}  = 2T^{1-s} \pi^{-s}
\left( 1 - 2^{-s} \right) \zeta (s)
\end{equation}
From this, note that
\begin{eqnarray}
T \sum_{\omega_n} 1 &=&  g(0) = 0 \nonumber \\
 T \sum_{\omega_n} \ln
|\omega_n| &=& - \left. \frac{dg}{ds} \right|_{s=0} = T \ln 2
\end{eqnarray}
Using these results in Eq.~(\ref{e1}) in taking the limit $d
\rightarrow 2$, we obtain
\begin{equation}
\chi_f =\left(\mbox{Tr} (T^a)^2 \right) \frac{T \ln 2}{\pi}
\end{equation}

{\em $\mathcal{A}_{\chi_f}$ at next to leading order in $1/N$:} We now turn to a computation of the diagrams of Fig.~\ref{fig:chiPsi}(b,c), these are the only contributions to $\chi_f$ at next to leading order. As will be clear below a full evaluation of these diagrams is not possible analytically. However we will be able to achieve our main goal, which is to extract the numerical value of the $1/N$ corrections to  the $N=\infty$ value of $\mathcal{A}_{\chi_f}$ computed above.

A direct evaluation of the two diagrams, Fig.~\ref{fig:chiPsi}(b,c) can be written in the form (including the $N=\infty$ result above),
\begin{widetext} 
\begin{eqnarray}
\label{eq:chiPsi1onN}
\chi_f &=& \left(\mbox{Tr} (T^a)^2 \right) \left\{ D_1 (0,0) +
\frac{1}{N} T \sum_{\epsilon_n} \int \frac{d^2 k}{4 \pi^2} \left[
\frac{D_3 (k, \epsilon_n)}{D_1 (k, \epsilon_n )} - \frac{D_4 (k,
\epsilon_n)}{D_2 (k, \epsilon_n) + (\epsilon_n^2/k^2) D_1 (k,
\epsilon_n)} \right] \right\}\nonumber \\
D_3 ({\bf k}, \epsilon_n) &\equiv&
 N_f T \sum_{\omega_n} \int \frac{d^2 q}{4 \pi^2} \mbox{Tr} \left[
2   G^3_\Psi({\bf q},\omega_n)
G_\Psi ({\bf q+k},\omega_n+\epsilon_n)+
G^2_\Psi({\bf q},\omega_n)
G^2_\Psi ({\bf q+k},\omega_n+\epsilon_n)
\right] \nonumber\\
D_4 ({\bf k}, \epsilon_n) &\equiv&
 N_f T \sum_{\omega_n} \int \frac{d^2 q}{4 \pi^2} \mbox{Tr} \left[
2   G^3_\Psi({\bf q},\omega_n)\sigma_i
G_\Psi ({\bf q+k},\omega_n+\epsilon_n)\sigma_j\right.\nonumber\\
&~&~~~~~~~~~~~~~~~~~~\left.
+~~G^2_\Psi({\bf q},\omega_n)\sigma_i
G^2_\Psi ({\bf q+k},\omega_n+\epsilon_n)\sigma_j
\right] (\delta_{ij}-\frac{k_ik_j}{k^2})\nonumber
\end{eqnarray}
\end{widetext} 
where the functions $D_{1,2}$ have been introduced (Section~\ref{sec:proptfin}) and computed (Appendix~\ref{app:D1D2}) before. Computations of $D_{3,4}$ are presented in Appendix~\ref{app:D3D4}. We note here that the dependence of $\mathcal{A}_{\chi_f}$ on $N_b$ enters through the $N_b$ dependence of $D_{1,2}$. A full numerical evaluation of $\mathcal{A}_{\chi_f}$ was carried out and was found to reproduce the results of Sec.~\ref{sec:univamp} up to three significant digits.

\section{\label{app:D3D4} Evaluation of $D_3$ and $D_4$}

In this Appendix we summarize our strategy to evaluate the functions $D_3$ and $D_4$ defined in  Eq.~(\ref{eq:chiPsi1onN}). The techniques are similar to those used to evaluate $D_{1,2}$ and we will simply quote the final answers,
\begin{widetext}
\begin{eqnarray}
D_3 ({\bf k}, \epsilon_n) &=&
 N T \sum_{\omega_n} \int \frac{d^2 q}{4 \pi^2} \mbox{Tr} \left[
2 \left(-i \omega_n + \vec{\sigma} \cdot {\bf q}\right)^{-3}
\left(-i (\omega_n + \epsilon_n) + \vec{\sigma} \cdot ({\bf q}+{\bf
k})\right)^{-1} \right. \nonumber \\
&~&~~~~~~~~~~~~~~~~~~~~~~~~~~~~~~~~+\left. \left(-i \omega_n +
\vec{\sigma} \cdot {\bf q}\right)^{-2} \left(-i (\omega_n +
\epsilon_n) + \vec{\sigma} \cdot ({\bf q}+{\bf k})\right)^{-2}
\right] \nonumber
\\
&=& \frac{N T}{2\pi} \int_0^1 dx \sum_{\omega_n} \Bigl[ -2 (x-1)
\omega_n^4-(x-1) (5 x \epsilon_n +\epsilon_n ) \omega_n^3\nonumber \\
&~&~~~~~~~~~~~-(x-1)
\left(2 x^2 k^2-x
   k^2-2 x^2 \epsilon_n^2+7 x \epsilon ^2\right) \omega_n^2 \nonumber\\
   &~&~~~~~~~~~~~-(x-1) \left(7 k^2 \epsilon_n
   x^3-3 \epsilon_n^3 x^2-8 k^2 \epsilon_n  x^2+3 \epsilon_n^3 x+3 k^2 \epsilon_n  x\right)
   \omega_n \nonumber \\
   &~&~~~~~~~~~~~-(x-1) \left((x-1) x^2 k^4+3 x^3 \epsilon_n^2 k^2-2 x^2 \epsilon_n^2 k^2-x^2
   \epsilon_n^4\right)
   \Bigr] \nonumber \\
   &~&~~~~~~~~~~~~~~~~~~~~~~~~~~\times\frac{(-1)}{((1-x) \omega_n^2 + x (\omega_n +
\epsilon_n)^2 +
  k^2 x(1-x))^3}.
\end{eqnarray}
This last lengthy expression was generated in \LaTeX\  by
Mathematica, and also translated to Fortran by Mathematica. The
integral over $x$ was evaluated numerically, and the frequency
summation was performed using the following expansion at large
$\omega_n$
\begin{equation}
- \frac{1}{\omega_n^2} - \frac{k^2 - \epsilon_n^2}{2 \omega_n^4} +
\frac{-k^4 +7 k^2 \epsilon_n^2 - 2 \epsilon_n^4 }{6 \omega_n^6}
\end{equation}
Similarly for $D_4$, we have
\begin{eqnarray}
D_4 ({\bf k}, \epsilon_n) &=&
 \left( \delta_{ij} - \frac{k_i k_j}{k^2} \right)
 N T \sum_{\omega_n} \int \frac{d^2 q}{4 \pi^2} \mbox{Tr} \left[
2 \left(-i \omega_n + \vec{\sigma} \cdot {\bf q}\right)^{-3}
\sigma_i  \left(-i (\omega_n + \epsilon_n) + \vec{\sigma} \cdot
({\bf q}+{\bf
k})\right)^{-1} \sigma_j \right. \nonumber \\
&~&~~~~~~~~~~~~~~~~~~~~~~~~~~~~~~~~+\left. \left(-i \omega_n +
\vec{\sigma} \cdot {\bf q}\right)^{-2} \sigma_i \left(-i (\omega_n +
\epsilon_n) + \vec{\sigma} \cdot ({\bf q}+{\bf k})\right)^{-2}
\sigma_j  \right] \nonumber
\\
&=& \frac{N T}{2\pi} \int_0^1 dx \sum_{\omega_n} \Bigl[ -(1-3 x)
(x-1) \omega_n^4-(x-1) \left(-10 \epsilon_n  x^2+3 \epsilon_n
x+\epsilon_n \right)
   \omega_n^3 \nonumber \\ &~&~~~~~-(x-1) \left(-6 k^2 x^3-8 \epsilon_n^2 x^3-k^2 x^2-7 \epsilon_n^2 x^2+8 k^2
   x+8 \epsilon_n^2 x\right) \omega_n^2 \nonumber \\
   &~&~~~~~-(x-1) \left(-6 k^2 \epsilon_n  x^4-8 \epsilon_n^3
   x^3+k^2 \epsilon_n  x^3+\epsilon_n^3 x^2+4 k^2 \epsilon_n  x^2+3 \epsilon_n^3 x+3 k^2
   \epsilon_n  x\right) \omega_n \nonumber \\
   &~&~~~~~-(x-1) \left(x^2 \left(x^3-4 x^2+6 x-3\right) k^4-3 x^4
   \epsilon_n^2 k^2+6 x^3 \epsilon_n^2 k^2-2 x^2 \epsilon_n^2 k^2-2 x^3 \epsilon_n^4+x^2
   \epsilon_n^4\right)
   \Bigr] \nonumber \\
   &~&~~~~~~~~~~~~\times \frac{(-1)}{((1-x) \omega_n^2 + x (\omega_n +
\epsilon_n)^2 +
  k^2 x(1-x))^3}.
\end{eqnarray}
The frequency summand has the following expansion at large
$\omega_n$
\begin{equation}
 \frac{-2 k^2 +  \epsilon_n^2}{2 \omega_n^4} + \frac{4 k^4 -17 k^2
\epsilon_n^2 - \epsilon_n^4 }{6 \omega_n^6}
\end{equation}
\end{widetext}



\end{document}